\documentclass[prd,aps,tightenlines,a4paper,12pt]{revtex4}

\usepackage{graphicx}
\usepackage{bm}

\usepackage{url}
\usepackage{amsmath}

\usepackage{natbib}

\newcommand{\OO}[1]{{\mathcal O}(c^{-#1})}

\newcommand{\muas}[0]{\hbox{\rm $\mu$as}}

\newcommand{\ve}[1]{\mbox{\boldmath$#1$}}

\begin{document}

\title{Using a coordinate-independent impact parameter in the post-post-Newtonian
light deflection formulas}

\author{Sergei A. \surname{Klioner}, Sven \surname{Zschocke}}

\affiliation{Lohrmann Observatory, Dresden Technical University,
Mommsenstr. 13, 01062 Dresden, Germany}

\begin{abstract}
\begin{center}
{\bf GAIA-CA-TN-LO-SK-008-1}

\medskip

\today

\end{center}

In \cite{report2} we have seen that among the post-post-Newtonian
terms in the light deflection formulas there are ``enhanced'' ones
that may become much larger than the other post-post-Newtonian terms.
It is demonstrated here that these ``enhanced'' terms result from an
inadequate choice of the impact parameter.  Introducing another impact
parameter, that can be considered as coordinate-independent, we
demonstrate that all ``enhanced'' terms disappear from the formulas.

\end{abstract}


\maketitle


\tableofcontents

\newpage

\section{Introduction}

In \cite{report1} an analytical post-post-Newtonian solution of light
propagation in the Schwarzschild metric generalized using PPN and
post-linear parameters has been presented. In \cite{report2} various
terms in the transformations between the units vectors $\ve{\sigma}$,
$\ve{n}$, and $\ve{k}$ characterizing light propagation have been
analytically estimated. These estimates reveal that in each
transformation ``peculiar'' or ``enhanced'' post-post-Newtonian terms
exist that can become much larger that the other ``regular''
post-post-Newtonian terms. In each case the sum of the ``regular''
post-post-Newtonian terms can be estimated as
$\frac{\displaystyle 15}{\displaystyle 4}\,\pi\,\frac{\displaystyle m^2}{\displaystyle d^2}$, 
$m$ being the Schwarzschild radius
of the central body and $d$ is the impact parameter.  The ``enhanced''
terms can be much larger (being, however, of order $m^2$).  Below we
demonstrate that the ``enhanced'' terms result from an inadequate choice
of parametrization of the light ray.  Indeed, one can demonstrate that
the ``enhanced'' terms disappear if the light deflection formulas are
expressed through a coordinate-independent impact parameter.

\section{Impact parameters}

The analytical solutions of the four basic transformations $\ve{k}$ to
$\ve{\sigma}$, $\ve{\sigma}$ to $\ve{n}$, $\ve{k}$ to $\ve{n}$, and
$\ve{\sigma}$ to $\ve{n}$ for stars and quasars, given in 
\citep{report1,report2}, are expressed through one of the two following
impact parameters:
\begin{eqnarray}
\ve{d}_{\sigma} &=& \ve{\sigma} \times \left( \ve{x} \times \ve{\sigma}\right),
\label{impact_10}
\\
\ve{d}_{\ } &=& \ve{k} \times \left( \ve{x}_0 \times \ve{k}\right) = \ve{k} \times \left( \ve{x} \times \ve{k}\right),
\label{impact_5}
\end{eqnarray}

\noindent
where $\ve{x}_0$ is the position of the source and $\ve{x}$ is the
position of the observer. The definitions of $\ve{\sigma}$ and
$\ve{k}$ are given in \citep{report1}.  The absolute values of these
impact vectors are denoted by $d=\left|\ve{d}\right|$ and
$d_{\sigma} = \left| \ve{d}_{\sigma} \right|$. 

In Section III.B of \citep{report1} the absolute value 
$d^{\;\prime} = \left|\ve{d}^{\;\prime}\right|$ of yet another
impact parameter
\begin{equation}
\ve{d}^{\;\prime} = \lim_{t \to -\infty} \ve{\sigma} \times \left( \ve{x}(t) \times \ve{\sigma}\right)
\label{impact_20}
\end{equation}
\noindent
has been introduced in order to compare the expression of total light
deflection derived there with the results found by other authors. Note that
here $\ve{x}(t)$ is the position of the photon at some arbitrary moment of time $t$.
For $\ve{d}^{\;\prime}$ one also has
\begin{equation}
\ve{d}^{\;\prime} = 
\lim_{t \to-\infty} {1\over c}\,\dot{\ve{x}}(t) \times \left( \ve{x}(t) \times {1\over c}\,\dot{\ve{x}}(t)\right) .
\label{d-prime-xdot}
\end{equation}
\noindent
Here one should take into account that $\ve{\sigma}=\lim\limits_{t \to
  -\infty}{1\over c}\,\dot{\ve{x}}(t)$, $\dot{\ve{x}}(t)$ being the
velocity of the photon at time $t$.  For a similar impact parameter
defined at $t\to+\infty$
\begin{equation}
\ve{d}^{\;\prime\prime} = 
\lim_{t \to +\infty} {1\over c}\,\dot{\ve{x}}(t) \times \left( \ve{x}(t) \times {1\over c}\,\dot{\ve{x}}(t)\right)=
\lim_{t \to+\infty} \ve{\nu} \times \left( \ve{x}(t) \times \ve{\nu}\right),
\label{d-prime-prime}
\end{equation}
\noindent
where $\ve{\nu}=\lim\limits_{t \to+\infty}{1\over c}\,\dot{\ve{x}}(t)$,
one has $\left|\ve{d}^{\;\prime}\right|
=\left|\ve{d}^{\;\prime\prime}\right|$. It is also clear that the
angle between $\ve{d}^{\;\prime}$ and $\ve{d}^{\;\prime\prime}$
is equal to the full light deflection. Since both $\ve{d}^{\;\prime}$
and $\ve{d}^{\;\prime\prime}$ resided at time-like infinity (since they 
are defined for $|t|\to\infty$) and since the gravitational
system under study is asymptotically flat, these parameters 
can be called coordinate-independent.

One can show that $d^{\;\prime}=d^{\;\prime\prime}$ coincides with the
impact parameter $D$ introduced, e.g., by Eq. (215) of Section 20 of
\cite{Chandrasekhar1983} in terms of full energy and angular momentum
of the photon. Indeed, in polar coordinates $(x,\varphi)$ the Chandrasekhar's
impact parameter $D=f(x)\,x^2\,\dot{\varphi}$, where $\lim\limits_{x\to\infty}f(x)=1$.
Clearly, $x^2\,\dot{\varphi}=|\dot{\ve{x}}(t)\times\ve{x}(t)|$ and it is obvious
that $d^{\;\prime}=d^{\;\prime\prime}=D$. Interestingly, this discussion
allows one to find an exact integral of the equations of motion for a photon in
Schwarzschild field. In the exact Schwarzschild solution used in \cite{report2}
in harmonic coordinates, Eq. (11) of \cite{report2} has an integral
\begin{equation}
\ve{D}={(1+a)^3\over 1-a}\,{1\over c}\,\dot{\ve{x}}(t)\times\ve{x}(t)={\rm const},
\end{equation}
\noindent
while for the parametrized post-post-Newtonian equations of motion given
by Eq. (24) of \cite{report1} one has
\begin{eqnarray}
\ve{D}&=&\exp\biggl(2(1+\gamma)\,a+\alpha\,\left(2\,(1-\beta)+\epsilon-2\gamma^2\right)\,a^2
\biggr)\,{1\over c}\,\dot{\ve{x}}(t)\times\ve{x}(t)
\nonumber\\
&=&\left(1+2(1+\gamma)\,a+\left(2(1+\gamma)^2+\alpha\,\left(2\,(1-\beta)+\epsilon-2\gamma^2\right)\right)\;a^2\right)\,{1\over c}\,\dot{\ve{x}}(t)\times\ve{x}(t)+\OO{6}
\nonumber\\
&=&{\rm const},
\label{D-ppN}
\end{eqnarray}
\noindent
where all the notations are as in the corresponding equations in
\cite{report1,report2}.  First line of (\ref{D-ppN}) represents an
exact integral of the [approximate] equation of motion (24) of
\cite{report2}.  In both cases the Chandrasekhar's $D$ is the absolute
value of $\ve{D}$ as given above.

The aim of this investigation is to express all the transformations
between vectors $\ve{\sigma}$, $\ve{n}$, and $\ve{k}$ using the
coordinate-independent impact vector $\ve{d}^{\;\prime}$. Therefore,
we need to have a relation between impact parameters (\ref{impact_5})
and (\ref{impact_10}), respectively, and (\ref{impact_20}). Relation
between $\ve{d}^{\;\prime}$ and $\ve{d}_\sigma$ can be derived 
using the post-Newtonian solution for light propagation given 
in Section III.A of \cite{report1} and has been partially
given by Eq. (43) in Section  III.B of \cite{report1}. One gets
\begin{equation}
\ve{d}^{\;\prime} = \ve{d}_{\sigma} \left( 1 + (1+\gamma) \;\frac{m}{d_{\sigma}^2}\;
\left(x + \ve{\sigma} \cdot \ve{x} \right) \right) + {\cal O} \left(m^2\right).
\label{impact_25}
\end{equation}
\noindent
Relation of $\ve{d}^{\;\prime}$ and $\ve{d}$ can be derived by
considering Eq. (31) in Section III.C of \cite{report1} 
in post-Newtonian approximation. Substituting this relation
into the definition of $\ve{d}_\sigma$ and considering
(\ref{impact_25}) one gets
\begin{equation}
\ve{d}^{\;\prime} = \ve{d} \left( 1 + \left(1 + \gamma \right) 
\;\frac{m}{d^2}\;
\frac{x + x_0}{R}\;\frac{R^2 - \left(x - x_0 \right)^2}{2\;R} \right) 
- (1 + \gamma)\;m\;\ve{k}\;\frac{x-x_0 +R}{R} 
+ {\cal O} \left(m^2\right).
\label{impact_30}
\end{equation}

Now let us consider the transformations between $\ve{\sigma}$,
$\ve{n}$, and $\ve{k}$.

\section{Transformation between $\ve{k}$ and $\ve{\sigma}$}

The transformation between $\ve{k}$ and $\ve{\sigma}$ is 
given by Eq. (29) of \cite{report2} or, retaining only  
``enhanced'' post-post-Newtonian terms, by Eqs. (35)--(36)
of \cite{report2}. These latter equations can be re-written 
as follows
\begin{eqnarray}
\ve{d}\,S\,
\left(1-S\,{1\over 2}\,(x+x_0)\left(1+{x_0-x\over R}\right)\right)
&=&\ve{d}^{\;\prime}\,S^{\;\prime} 
\nonumber\\
&&
+ (1+\gamma)^2\;m^2\;\frac{\left(x - x_0 + R\right)^2}{\left| \ve{x} \times \ve{x}_0 \right|^2} \;\ve{k}
+{\cal O}\left({m^3}\right),
\label{S-rewriting}
\end{eqnarray}
\noindent
where $S$ is defined by Eq. (36) of \cite{report2}, and $S^{\;\prime}$
has the same functional form, but with $d$ replaced by $d^{\;\prime}$:
\begin{equation}
S^{\;\prime}=(1+\gamma)\,{m\over d^{\;\prime2}}\,\left(1-{x_0-x\over R}\right)\,.
\label{S-prime}
\end{equation}
\noindent
The term in (\ref{S-rewriting}) proportional to $\ve{k}$ is of type ``scaling''
as explained in Section V.B of \cite{report2} and plays no role for the light deflection
(its absolute value can be estimated as $8\,m^2/d^2$). Therefore, the transformation
between $\ve{k}$ and $\ve{\sigma}$ can be finally written as
\begin{equation}
\ve{\sigma} = \ve{k}\,+\ve{d}^{\;\prime}\,S^{\;\prime}
+ {\cal O}\left(\frac{m^2}{d^2}\right) + {\cal O}\left({m^3}\right)\,.
\label{sigma-k-postNewtonian_10}
\end{equation}
\noindent
Note that the terms of order $m^2/d^2$ in (\ref{sigma-k-postNewtonian_10})
have the same upper estimate given by Eq. (33) of \cite{report2}.

\section{Transformation between $\ve{\sigma}$ and $\ve{n}$}

The transformation between $\ve{\sigma}$ and $\ve{n}$ is given by 
Eq. (37) of \cite{report2} or, retaining only ``enhanced'' post-post-Newtonian
terms, by Eqs. (43)--(44) of \cite{report2}. The latter equations can be
rewritten in terms of $\ve{d}^{\;\prime}$ as 
\begin{eqnarray}
\ve{d}\,T\,\left(1+T\,x\,
{R+x_0-x\over R+x_0+x} \right) 
&=&\ve{d}^{\;\prime}\,T^\prime
- (1+\gamma)^2 \;\frac{m^2}{d^2}\;\left(1 + \frac{\ve{k} \cdot \ve{x}}{x}\right) \frac{x - x_0 + R}{R}\;\ve{k}
\nonumber\\[8pt]
&&+\ve{\varphi}_{\rm corr}+ {\cal O}\left(m^3\right)
\label{T-rewriting}
\end{eqnarray}
\noindent
where $T$ is defined by Eq. (44) of \cite{report2} and 
\begin{eqnarray}
T^{\prime} &=& -(1 + \gamma)\,\frac{m}{d^{\;\prime2}}\,
\left(1 + \frac{\ve{k} \cdot \ve{x}}{x}\right)\,,
\label{T-prime}
\\
\ve{\varphi}_{\rm corr} &=& -(1 + \gamma)^2\;\frac{m^2}{d^2}
{x-x_0+R\over R}\,
\frac{\ve{d}}{x}\,.
\label{phi_corr_5}
\end{eqnarray}
\noindent
In \cite{report2} the sum of all ``regular'' post-post-Newtonian terms
in the transformation from $\ve{\sigma}$ to $\ve{n}$ was denoted by
$\ve{\varphi}_{\rm ppN}$ and estimated to have absolute value less than
$\frac{\displaystyle 15}{\displaystyle 4}\,\pi\,\frac{\displaystyle m^2}{\displaystyle d^2}$ 
(see Section V.C of \cite{report2}). One can also demonstrate (see Appendix \ref{appendix-proof-of-phi-corr}) that
\begin{equation}
\left|\,\ve{\varphi}_{\rm ppN} + \ve{\varphi}_{\rm corr}\,\right| \le \frac{15}{4}\;\pi\;\frac{m^2}{d^2}\,.
\label{phi_corr_10}
\end{equation}
\noindent
The term in (\ref{T-rewriting}) 
proportional to vector $\ve{k}$ is again a ``scaling'' term
and can be omitted since it does not influence the direction of $\ve{n}$. 
Again its absolute value is of order $m^2/d^2$.
Finally, the transformation between $\ve{\sigma}$ and $\ve{n}$ can be 
written as
\begin{equation}
\ve{n}=\ve{\sigma}+\ve{d}^{\;\prime}\,T^{\;\prime} 
+ {\cal O} \left(\frac{m^2}{d^2}\right) + {\cal O} \left(m^3\right)\,.
\label{n-sigma-postNewtonian_10}
\end{equation}
\noindent
Again  the terms of order $m^2/d^2$ in (\ref{n-sigma-postNewtonian_10})
have the same upper estimate given by Eq. (41) of \cite{report2}.

\section{Transformation $\ve{k}$ to $\ve{n}$}

The transformation between $\ve{n}$ and $\ve{k}$ is given by 
Eq. (45) of \cite{report2} or, retaining only ``enhanced'' post-post-Newtonian
terms, by Eqs. (52)--(53) of \cite{report2}. The latter equations can be
rewritten in terms of $\ve{d}^{\;\prime}$ as 
\begin{eqnarray}
\ve{d}\,P\,\left(1+P\,x\,{x_0+x\over R}\right) 
&=&
\ve{d}^{\;\prime}\,P^{\;\prime}
\nonumber\\
&& 
- (1+\gamma)^2\;\frac{m^2}{d^2}\;\frac{x - x_0 + R}{R}\left(\frac{x_0 - x}{R} + \frac{\ve{k}\cdot \ve{x}}{x}\right)\ve{k}
+ {\cal O}\left(m^3\right)\,,
\label{P-rewriting}
\end{eqnarray}
\noindent
where $P$ is defined by Eq. (53) of \cite{report2}, and $P^\prime$ has the
same functional form as $P$, but with $d$ replaced by $d^{\;\prime}$: 
\begin{eqnarray}
P^\prime &=&-(1+\gamma)\,{m\over d^{\prime2}}\,
\left({x_0-x\over R}+{\ve{k}\cdot\ve{x}\over x}\right)\,.
\label{P-sso}
\end{eqnarray}
\noindent
The term in (\ref{P-rewriting}) proportional to vector $\ve{k}$ 
is again ``scaling''
and plays no role for the light deflection.
Finally, 
the transformation between
$\ve{n}$ and $\ve{k}$ can be written as
\begin{equation}
\ve{n}=\ve{k}+\ve{d}^{\;\prime}\,P^{\;\prime} 
+ {\cal O}\left(\frac{m^2}{d^2}\right)\, + {\cal O}\left(m^3\right)\,.
\label{n_85-postNewtonian_10}
\end{equation}

\noindent
Also in this equation terms of order $m^2/d^2$ 
have the same upper estimate given by Eq. (50) of \cite{report2}.

\section{Transformation $\ve{\sigma}$ to $\ve{n}$ for stars and quasars}

Finally, let us consider transformation from $\ve{\sigma}$ to $\ve{n}$
for stars and quasars. This transformation is given by 
Eq. (58) of \cite{report2} or, retaining only ``enhanced'' post-post-Newtonian
terms by Eqs. (62)--(63), of \cite{report2}. The latter equations can be
rewritten in terms of $\ve{d}^{\;\prime}$ as 
\begin{equation}
\ve{d}_\sigma\,Q\,(1+Q\,x)=\ve{d}^{\;\prime}\,Q^\prime
+{\cal O}\left(m^3\right)\,,
\label{Q-rewriting}
\end{equation}
\noindent
where $Q$ is defined by Eq. (63) of \cite{report2}, and $Q^\prime$ has the same functional
form as $Q$ with $d_\sigma$ replaced by $d^{\;\prime}$:
\begin{equation}
Q^\prime = -(1+\gamma)\,{m\over d^{\;\prime2}}\,\left(1+{\ve{\sigma}\cdot\ve{x}\over x}\right)\,.
\label{Q-stars}
\end{equation}
\noindent
Therefore, transformation from $\ve{\sigma}$ to $\ve{n}$ can be written as
\begin{equation}
\ve{n} = \ve{\sigma}+\ve{d}^{\;\prime}\,Q^{\;\prime} + {\cal O}\left(\frac{m^2}{d_{\sigma}^2}\right) 
+ {\cal O}\left(m^3\right)\,.
\label{sigma-n-stars-postNewtonian_10}
\end{equation}

\section{Summary}

Thus, we have demonstrated that the source of the ``enhanced''
post-post-Newtonian terms discussed in \cite{report2} is an inadequate
choice of the impact parameter. Using the coordinate-independent
coordinate parameter $\ve{d}^{\;\prime}$ discussed above one can
eliminate the ``enhanced'' post-post-Newtonian terms from the
analytical formulas. Above we have demonstrated that the four relevant
transformations between vectors $\ve{\sigma}$, $\ve{n}$, and $\ve{k}$
can be expressed by Eqs.  (\ref{sigma-k-postNewtonian_10}),
(\ref{n-sigma-postNewtonian_10}), (\ref{n_85-postNewtonian_10}), and
(\ref{sigma-n-stars-postNewtonian_10}). In all these formulas
the omitted post-post-Newtonian terms can be estimated to be less
than $\frac{\displaystyle 15}{\displaystyle 4}\,\pi\,\frac{\displaystyle m^2}{\displaystyle d^2}$ 
and, therefore, the formulas guarantee numerical accuracy of 1 \muas\ in each case as long as observations
of sources within 5 angular radii from the Sun are not considered.

Although this investigation elucidates the origin of the ``enhanced''
post-post-Newtonian terms, the results given by Eqs.
(\ref{sigma-k-postNewtonian_10}), (\ref{n-sigma-postNewtonian_10}),
(\ref{n_85-postNewtonian_10}), and
(\ref{sigma-n-stars-postNewtonian_10}) are strictly equivalent to
those derived in \cite{report2} and are not necessarily more
convenient from the computational point of view than the formulas
in terms of $\ve{d}$ and $\ve{d}_\sigma$.

\acknowledgments

This work was partially supported by the BMWi grants 50\,QG\,0601
and 50\,QG\,0901
awarded by the Deutsche Zentrum f\"ur Luft- und Raumfahrt e.V. (DLR).


\appendix

\section{Proof of (\ref{phi_corr_10})}
\label{appendix-proof-of-phi-corr}

Eq. (C3) of \cite{report1} demonstrates that
\begin{eqnarray}
\left|\,\ve{\varphi}_{\rm ppN}\right| &=&{m^2\over d^2}\,{1\over 4}\,f_7,
\label{varphi-ppN-est}
\\
f_7&=&\left|
16\,\sin\Psi
+\sin\Psi\,\cos\Psi
-2\,\sin^3\Psi\,\cos\Psi
-15\,(\pi-\Psi)
\right|\,,
\label{ppN-estimate}
\end{eqnarray}
\noindent
where $\Psi$ is the angle between vectors $\ve{k}$ and $\ve{n}$
($0\le\Psi\le\pi$).  For $\ve{\varphi}_{\rm corr}$ defined by
(\ref{phi_corr_5}) one can write (as usual we assume $\gamma=1$ here)
\begin{eqnarray}
|\ve{\varphi}_{\rm corr}| &=& 4\,\frac{m^2}{d^2}\,{d\over x}\,{x-x_0+R\over R}
=4\,\frac{m^2}{d^2}\,{d\over x}\,f_3,
\label{phi-corr-est1}
\\
f_3&=&{1-z\over\sqrt{1+z^2-2z\cos\Phi}}+1,
\label{f_3}
\end{eqnarray}
\noindent
where $\Phi=\delta(\ve{x},\ve{x}_0)$ is the angle between $\ve{x}$ and
$\ve{x}_0$, and $z=x_0/x$. Function $f_3$ has been already considered
in \cite{report2} (see Eq. (B2) of that report) and found to be less
than 2 (it is obvious that $f_3\ge0$).  Therefore, one has
\begin{equation}
|\ve{\varphi}_{\rm corr}| \le 8\;\frac{m^2}{d^2}\,\sin{\Psi}.
\label{phi-corr-est2}
\end{equation}
\noindent
Finally, combining 
(\ref{varphi-ppN-est}) and 
(\ref{phi-corr-est2}) one gets
\begin{equation}
|\ve{\varphi}_{\rm ppN}+\ve{\varphi}_{\rm corr}|\le
|\ve{\varphi}_{\rm ppN}|+|\ve{\varphi}_{\rm corr}|\le
\frac{m^2}{d^2}\,\left({1\over 4}\,f_7+8\,\sin\Psi\right).
\label{ppN+phi-2}
\end{equation}
\noindent
It is not difficult to see that 
\begin{equation}
f_7^{\rm corr}={1\over 4}\,f_7+8\,\sin\Psi\le {15\,\pi\over 4},
\label{f7}
\end{equation}
\noindent
and this immediately gives (\ref{phi_corr_10}).

\end{document}